\documentclass[journal]{IEEEtran}
\usepackage{float}
\usepackage{graphicx}
\usepackage{subfigure} 
\usepackage{epstopdf}
\usepackage{color}
\usepackage{amssymb}
\usepackage{amsmath}
\usepackage{algorithm}
\usepackage{algorithmic}
\usepackage{bm}
\usepackage[colorlinks]{hyperref}
\begin{document}
	
	\title{ \vspace{-7mm} \Large Towards Intelligent Antenna Positioning: Leveraging DRL for FAS-Aided ISAC Systems}
	\author{Shunxing Yang, Junteng Yao, Jie Tang,  Tuo Wu,  Maged Elkashlan, Chau Yuen, \emph{Fellow}, \emph{IEEE},  \\ M\'{e}rouane Debbah,  \emph{Fellow},  \emph{IEEE}, Hyundong Shin, \emph{Fellow}, \emph{IEEE}, and  Matthew Valenti, \emph{Fellow}, \emph{IEEE}
		\vspace{-10mm}
		\thanks{\emph{(Corresponding author: Tuo Wu.)}}
		\thanks{S. Yang is with China Telecom Research Institute, Guangzhou 510000, China (E-mail: $\rm yangsx@chinatelecom.cn$). J. Yao is with the faculty of Electrical Engineering and Computer Science, Ningbo University, Ningbo 315211, China (E-mail: $\rm yaojunteng@nbu.edu.cn$). J. Tang is with the School of Electronic and Information Engineering, South China University of Technology, Guangzhou 510640, China (E-mail: $\rm eejtang@scut.edu.cn$).  T. Wu and C. Yuen are with the School of Electrical and Electronic Engineering, Nanyang Technological University 639798, Singapore (E-mail: $\rm \{tuo.wu, chau.yuen\}@ntu.edu.sg$). M. Elkashlan is with the School of Electronic Engineering and Computer Science at Queen Mary University of London, London E1 4NS, U.K. (E-mail: $\rm maged.elkashlan@qmul.ac.uk$).  M. Debbah is with  KU 6G Research Center, Department of Computer and Information Engineering, Khalifa University, Abu Dhabi 127788, UAE (E-mail: $\rm merouane.debbah@ku.ac.ae$). H. Shin is with the Department of Electronic Engineering, Kyung Hee University, Yongin-si, Gyeonggi-do 17104, Korea (E-mail: $\rm hshin@khu.ac.kr$). M. Valenti is with the Lane Department of Computer Science and Electrical Engineering, West Virginia University, Morgantown, USA (E-mail: $\rm valenti@ieee.org$).} 
	}
	
	\maketitle
	\thispagestyle{empty} 
	\begin{abstract}
		Fluid antenna systems (FAS) enable dynamic antenna positioning, offering new opportunities to enhance integrated sensing and communication (ISAC) performance. However, existing studies primarily focus on communication enhancement or single-target sensing, leaving multi-target scenarios underexplored. Additionally, the joint optimization of beamforming and antenna positions poses a highly non-convex problem, with traditional methods becoming impractical as the number of fluid antennas increases. To address these challenges, this letter proposes a block coordinate descent (BCD) framework integrated with a deep reinforcement learning (DRL)-based approach for intelligent antenna positioning. By leveraging the deep deterministic policy gradient (DDPG) algorithm, the proposed framework efficiently balances sensing and communication performance. Simulation results demonstrate the scalability and effectiveness of the proposed approach.
	\end{abstract}
	
	\vspace{-5mm}
	\section{Introduction}
	\IEEEPARstart{F}{luid} antenna systems (FAS), capable of dynamically repositioning antennas within a defined area, have gained significant attention for their potential to enhance wireless system performance through additional spatial flexibility \cite{WKN242}. This adaptability aligns well with the demands of integrated sensing and communication (ISAC), a key innovation in sixth-generation (6G) wireless networks that combines communication and sensing functionalities over shared spectrum resources \cite{SLu24}.
	
	Traditional ISAC systems with fixed-position antennas (FPAs) often face fundamental challenges in balancing communication and sensing performance, particularly when multiple sensing targets are involved \cite{JYao24}. The fixed antenna positions limit the system's flexibility in spatial resource utilization, making it difficult to simultaneously optimize multi-target sensing beamforming and communication performance. While existing studies have proposed various optimization methods to design beamforming vectors, these approaches are constrained by the fixed antenna positions, leading to suboptimal system performance \cite{TWu2024}.
	
By contrast, fluid antennas (FAs) offer a transformative solution for ISAC by enabling dynamic antenna positioning at both the base station (BS) and communication users \cite{KKWong23, WMa23, WMei2024}. This flexibility can potentially address more demanding scenarios, including multi-target sensing. However, most existing works on FAs, such as \cite{LZhou24}, focus on single-target or purely communication-driven cases, leaving open whether FAs retain their advantages when sensing multiple targets. Meanwhile, the studies in \cite{JZou2024,CWang24} address FAs by selecting which ports to activate rather than continuously optimizing antenna locations, thus overlooking a critical dimension of FAS design. Furthermore, conventional alternating optimization methods \cite{LZhou24} become prohibitively complex when scaling up the number of sensing targets or FAs, underscoring the need for more efficient approaches.
	
To bridge these gaps, this letter investigates an ISAC system in which a dual-functional BS simultaneously communicates with a user terminal (UT) and performs radar sensing for multiple targets, emphasizing how to balance these two objectives. We propose a block coordinate descent (BCD) framework that alternately optimizes beamforming vectors and antenna positions. Central to this framework is a deep reinforcement learning (DRL)-based solution, built upon the deep deterministic policy gradient (DDPG) algorithm, which directly optimizes fluid antenna locations without relying on traditional iterative schemes. Compared with \cite{LZhou24}, our system design explicitly accommodates multiple sensing users; in contrast to \cite{JZou2024,CWang24}, we incorporate continuous position selection rather than merely activating discrete ports. Simulation results show that the proposed method not only achieves superior performance by leveraging FAS in multi-target sensing but also demonstrates remarkable scalability—extending naturally to more complex setups with additional users and sensing targets while enabling real-time decision-making.

	\vspace{-2mm}
	\section{ System Model and Problem Formulation} 
	\subsection{System Model}
	Consider an ISAC system comprising a dual-functional BS with $N$ FAs,  $1$ UT with a single FA \footnote{For ease of exposition, we consider one UT, e.g.,   $M=1$, as generalizing to $M>1$ is straightforward but complicates the notation. Later we give simulation results for $M>1$ to provide more insights.}, and $K$ sensing targets. The BS simultaneously communicates with the UT and performs radar sensing directed towards $K$ sensing targets. The FAs are connected to radio frequency (RF) chains via integrated waveguides or flexible cables, allowing free switching (or movement) within defined regions, denoted as $\mathcal{A}_t$ for the BS and $\mathcal{A}_r$ for the UT, respectively. 
	
	The location of the $n$-th FA of the BS is defined using a two-dimensional Cartesian coordinate model and can be expressed as $\mathbf{p}_{n}=[x^{(t)}_n,y_n^{(t)}]^{\mathrm T}, n\in\mathcal{N}=\{1,\dots,N\}$. The collective locations of the BS's FAs are represented as $\mathbf{\overline{p}} = [\mathbf{p}_1, \mathbf{p}_2, \dots, \mathbf{p}_N] \in \mathbb{R}^{2 \times N}$. Similarly, the location of the UT's single FA is denoted by $\mathbf{q} = [x^{(r)}, y^{(r)}]^{\mathrm T}$.
	
	We define the beamforming vector of the BS as $\mathbf{u} \in \mathbb{C}^{N \times 1}$, and denote $s$ as the transmit signal, used for both communication and radar sensing, with $\mathbb{E}[|s|^2]=1$. Consequently, the received signal at the UT can be expressed as
	\begin{equation}
		\mathbf{y}(\mathbf{\overline{p}}, \mathbf{q}) = \mathbf{f}(\mathbf{\overline{p}}, \mathbf{q}) \mathbf{u} s + z,
	\end{equation}
	where $\mathbf{f}(\mathbf{\overline{p}}, \mathbf{q}) \in \mathbb{C}^{1 \times N}$ represents the channel vector from the BS to the UT, and $z \sim \mathcal{CN}(0, \sigma^2)$ denotes the additive white Gaussian noise at the UT.
 \vspace{-5mm}
	\subsection{Channel Model} 
	We assume that the size of the `moving' region for the FAs is significantly smaller than the distance between the transmitter and the UT, thereby  adopting the far-field model in this letter \cite{WMa23}. Despite the mobility of the FAs, the angles of arrival (AoA) and angles of departure (AoD) remain constant for each propagation path. The variation in signal propagation distance for the $n$-th antenna at the BS relative to the reference origin on the ${d}$-th transmit path is expressed as 
	\begin{equation}
		\omega^{(t)}_{d}(\mathbf{p}_n)=x^{(t)}_{d}\sin\theta^{(t)}_{d} \cos\psi^{(t)}_{d}+y^{(t)}_{d}\cos\theta^{(t)}_{d},
	\end{equation}
	where $\sin\theta^{(t)}_{d} \in [0, \pi]$ and $\psi^{(t)}_{d} \in [0, \pi]$ represent the elevation and azimuth AoDs of the $d$-th path ($d \in \{1, \dots, D\}$) with respect to the boresight of the FAS array. Also, $D$ denotes the number of transmit paths from BS to UT. With the wavelength $\lambda$, the transmit response vector for the $n$-th FA is given by
	\begin{equation}\label{eq3}
 \mathbf{e}(\mathbf{p}_n) \triangleq \left [e^{j \frac{2\pi\omega^{(t)}_{1}(\mathbf{p}_n)}{\lambda}}, \cdots, e^{j \frac{2\pi\omega^{(t)}_{D}(\mathbf{p}_n)}{\lambda}}\right]^{\mathrm T} \in \mathbb{C}^{D\times 1}, n\in\mathcal{N}.
	\end{equation}
	Accordingly, the transmit response matrix from the BS to UT can be expressed as
	\begin{align}
		\mathbf{E(\overline{p})}  
		\triangleq \left [\mathbf{e}(\mathbf{p}_1),\mathbf{e}(\mathbf{p}_2), \cdots, \mathbf{e}(\mathbf{p}_N)\right]^{\mathrm T} \in \mathbb{C}^{D \times N}.
	\end{align}
	Similarly, the transmit response matrix from the BS to the $k$-th target is
	\begin{align}
 \hat{\mathbf{E}}(\overline{\mathbf{p}}^{(k)}) \triangleq \left [\hat{\mathbf{e}}(\mathbf{p}^{(k)}_1),\hat{\mathbf{e}}(\mathbf{p}^{(k)}_2), \dots, \hat{\mathbf{e}}(\mathbf{p}^{(k)}_N)\right]^{\mathrm T} \in \mathbb{C}^{P^{(k)}  \times N},
	\end{align}
	where $\theta^{(k)}_{p} \in [0, \pi]$ and $\psi^{(k)}_{p} \in [0, \pi]$ in $\hat{\mathbf{e}}(\mathbf{p}_n^{(k)}), n\in\mathcal{N}$, whose expression is similar with \eqref{eq3}, represent the elevation and azimuth AoDs of the $p^{(k)}$-th path, $P^{(k)}$ denotes the number of transmit paths from the BS to the $k$-th target.
	
	For the $i$-th receive path from the BS to UT, the difference in propagation distance between the single receive FA at the UT and its reference point is
	\begin{equation}
		\hspace{-5mm}\rho_i(\mathbf{q})=x^{(r)} \sin\theta^{(r)}_i \cos\psi^{(r)}_i +y^{(r)} \cos\theta^{(r)}_i , ~i \in \{1, \dots, I\}, \nonumber
	\end{equation}
	where $\theta^{(r)}_i\in [0, \pi]$ and $\psi^{(r)}_i\in [0, \pi]$ are the elevation and azimuth  AoAs  at the UT, respectively. The receive response vector for the fluid antenna at the UT is then given by
	\begin{equation}
		\mathbf{f({q})} \triangleq \left [e^{j \frac{2\pi}{\lambda}\rho_1(\mathbf{q})},\cdots,e^{j \frac{2\pi}{\lambda}\rho_I(\mathbf{q})}\right]^{\mathrm T} \in \mathbb{C}^{I \times 1},
	\end{equation}
	where $I$ represents the number of receive paths of the UT.

	Furthermore, we define the path response matrix $\bm{\Sigma}\in\mathbb{C}^{{K}\times {I}}$ as the responses of the transmit and receive paths between the BS and the UT. Therefore, the channel between the BS and the UT can be written as 
	\begin{equation}
		\mathbf{f(\overline {p}, {q})}=\mathbf{f}^{\dagger}\mathbf{(q)}\mathbf{\Sigma}\mathbf{E(\overline{p})}.
	\end{equation}
	where $\dagger$ denotes the  complex transpose. In the proposed ISAC   system,  the BS communicates with the receiever. Hence, the communication rate is considered as the performance metric, which is given by
	\begin{equation}
		R=\log_2\left(1+\frac{\mathbf{f(\overline {p}, {q})}\mathbf{U}\mathbf{f}^{\dagger}\mathbf{(\overline {p}, {q})} }{\sigma^{2}}\right),
	\end{equation}
	where $\mathbf{U}=\mathbf{u}\mathbf{u}^{\dagger}$.
	
	Parallel to this, the BS also employs the beamforming technique to enhance the sensing function. This approach is designed to direct a strong beampattern toward potential targets, thereby increasing the radar signal-to-noise ratio (SNR) and ultimately improving sensing performance. Consequently, we quantify the sensing performance by the sensing gain for the $k$-th sensing target, which is given by ${  \varpi}(\overline{\mathbf{p}}^{(k)})=\mathrm{Tr}\left(\hat{\mathbf{E}}(\overline{\mathbf{p}}^{(k)})  \mathbf{U}\hat{\mathbf{E}}^{\dagger}(\overline{\mathbf{p}}^{(k)}) \right)$, where $k\in\mathcal{K}=\{1,\dots,K\}$.
	
 \vspace{-3mm}
	\subsection{Problem Formulation}  
	We aim to maximize the communication rate while adhering to the constraints on the transmit power of the BS  and the sensing   gain. The optimizing variables include the transmit beamforming matrix $\mathbf{U}$, the locations of the transmit FAs $\mathbf{\overline{p}}$, and the location of the receive FA $\mathbf{q}$. Thus, the optimization problem can be formulated as
	\begin{subequations}\label{eq11}
		\begin{align}
			\max\limits_{\mathbf{\overline{p}},\mathbf{{q}},\mathbf{U}\succeq\mathbf{0}} \quad &R\label{07a}\\
			\mathrm{s.t.} \quad \ &\mathbf{\overline{p}} \in \mathcal{A}_t, \label{07b}\\
			&\mathbf{{q}} \in \mathcal{A}_r, \label{07c}\\
			&||\mathbf{p}_\alpha-\mathbf{p}_\beta||_2\geq D_s,~\alpha,\beta\in\mathcal{N},~\alpha\neq \beta, \label{07d}\\
			&\mathrm{Tr}(\mathbf{U}) \leq P_{\max}, \label{07e}\\
			&{\varpi}(\overline{\mathbf{p}}^{(k)})\geq\Gamma,~k\in\mathcal{K}, \label{07f}\\
			&\mathrm{rank}(\mathbf{U})=1, \label{07g}
		\end{align}
	\end{subequations}
	where \eqref{07d} is the minimum distance requirement between the antennas in the transmit region to avoid coupling, {and $D_s$ is the predefined minimum distance between the transmit antennas}; \eqref{07e} denotes the maximum transmit power constraint of the BS; \eqref{07f} represents the sensing beampattern gain requirement, {and $\Gamma$ is the predefined sensing beampattern gain}; \eqref{07g} is the rank-one constraint of $\mathbf{U}$ due to the lack of  spatial multiplexing. However, due to the highly non-convex objective function \eqref{07a}, constraints \eqref{07d}, \eqref{07f}, and \eqref{07g}, solving \eqref{eq11} becomes exceedingly challenging. To address this issue, we employ an DRL-BCD algorithm, whose details are given in the following section.
	\vspace{-3mm}
	\section{DRL-BCD Algorithm}  
	To solve the optimization problem in \eqref{eq11}, we propose a DRL-BCD algorithm that alternately optimizes the transmit covariance matrix and antenna positions. The algorithm decomposes \eqref{eq11} into two sub-problems: a convex optimization problem for the transmit covariance matrix and a deep reinforcement learning problem for the antenna positions. \vspace{-3mm}
	\subsection{Optimization of Transmit Covariance Matrix}  
	For fixed antenna locations $\mathbf{\overline{p}}$ and $\mathbf{q}$, by temporarily relaxing the constraint \eqref{07g}, Problem \eqref{eq11} can be reformulated as
	\begin{subequations}\label{eq12}
		\begin{align}
			\max\limits_{\mathbf{U}\succeq\mathbf{0}} \quad &R \label{eq12a}\\
			\mathrm{s.t.}\quad  & \eqref{07e},\eqref{07f}.
		\end{align}
	\end{subequations}
	The objective function is concave with respect to $\mathbf{U}$, and constraints \eqref{07e} and \eqref{07f} are linear, making \eqref{eq12} a convex optimization problem. We solve it efficiently using the CVX toolbox \cite{CVX}. If the obtained solution $\mathbf{U}$ has $\mathrm{rank}(\mathbf{U})>1$, we employ Gaussian randomization to reconstruct a rank-one solution.
 \vspace{-3mm}
	\subsection{Antenna Position Optimization via DDPG} 
	Given the optimized transmit covariance matrix $\mathbf{U}$, we address the antenna position optimization sub-problem. Due to its non-convex nature and continuous state-action space, we formulate it as a Markov decision process (MDP) and solve it using the deep deterministic policy gradient (DDPG) algorithm. The DDPG framework employs an actor-critic architecture, where the actor learns a deterministic policy that maps states to actions, while the critic evaluates action quality through Q-value estimation.
	\subsubsection{MDP Formulation}
	The MDP framework consists of three fundamental components: \textbf{state space}, \textbf{action space}, \textbf{reward function}, detailed as follows.
	\paragraph{State Space} 
	To capture the complete system dynamics, we design a comprehensive state space that incorporates both spatial configuration and beamforming characteristics. The system state $s_t \in \mathbb{R}^{2(N+1)+3}$ at time step $t$ integrates position information and key beamforming features. The spatial configuration is represented by the coordinates of all BS antennas $[x^{(t)}_1,y^{(t)}_1,\cdots,x^{(t)}_N,y^{(t)}_N]$ and the UT's fluid antenna position $[x^{(r)},y^{(r)}]$. To characterize the beamforming status, we extract three essential features from the beamforming matrix $\mathbf{U}$: the trace $\mathrm{tr}(\mathbf{U})$ quantifying total power consumption, the maximum eigenvalue $\lambda_{\max}(\mathbf{U})$ indicating the dominant transmission direction, and the mean eigenvalue $\bar{\lambda}(\mathbf{U})$ representing the average power distribution \footnote{These carefully selected beamforming features provide the DDPG agent with crucial information about the current communication and sensing state, enabling informed decision-making in antenna position optimization while maintaining effective coordination with beamforming optimization.}.
	
	\paragraph{Action Space}
	Building upon the state representation, we define the action space to control the movement of both BS and UT antennas. At each time step $t$, the action $a_t \in \mathbb{R}^{2(N+1)}$ represents the incremental position adjustments for all fluid antennas, formulated as
	\begin{equation}\label{eq13}
		a_t = [\Delta x_t^1,\Delta y_t^1,\dots,\Delta x_t^N,\Delta y_t^N,\Delta x_r,\Delta y_r],
	\end{equation}
	where $(\Delta x_t^n,\Delta y_t^n)$ denotes the position adjustment for the BS's $n$-th FA, and $(\Delta x_r,\Delta y_r)$ represents the movement of the UT's FA. To ensure practical antenna movements and system stability, each component of the action vector is bounded within the interval $\left[-{\frac{A}{2}},{\frac{A}{2}}  \right] \times \left[-{\frac{A}{2}},{\frac{A}{2}}  \right]$, where $A=4\lambda$ defines  maximum allowable displacement in terms of wavelength $\lambda$.  
	\paragraph{Reward Function}
	To guide the learning process towards optimal FA configurations that balance multiple system constranits, we design a comprehensive reward function. At each time step $t$, the reward $r_t$ is computed as 
	\begin{align} \label{eq14}   
	&r_t =  R(s_t,a_t) - \alpha_1 \sum_{m=1}^M \bigg(\max(0,{\varpi}(\overline{\mathbf{p}}^{(m)})-\Gamma \bigg) \nonumber\\
		&-  \alpha_2 \max(0,P_{\max}-\mathrm{Tr}(\mathbf{U})) -   \alpha_3 \frac{1}{N+1}\sum_{i=1}^{N+1}\|\Delta p_i\|_2, 
	\end{align}
	where $R(s_t,a_t)$ represents the communication rate under the current state and action pair. The second term, weighted by $\alpha_1$, penalizes violations of the required sensing gain constraint, ensuring that the sensing gain ${\varpi}(\overline{\mathbf{p}}^{(m)})$ for each target $m$ remains greater than $\Gamma$. The third term, with weight $\alpha_2$, enforces the total power constraint by penalizing any excess power consumption above the threshold $P_{\max}$.  The final term, weighted by $\alpha_3$, introduces a movement penalty  to reach the objective that is proportional to the average Euclidean distance of antenna movements, where $\|\Delta p_n\|_2 = \sqrt{(\Delta x_t^n)^2 + (\Delta y_t^n)^2}$ for BS antennas $(n=1,\dots,N)$ and $\|\Delta p_{N+1}\|_2 = \sqrt{(\Delta x_r)^2 + (\Delta y_r)^2}$ for the UT's FA. This reward function enables the DDPG agent to learn a policy that jointly optimizes communication performance while satisfying sensing, and power constraints.
	\begin{figure}[t]\vspace{-3mm}
		\centering
		\includegraphics[width=3.3in]{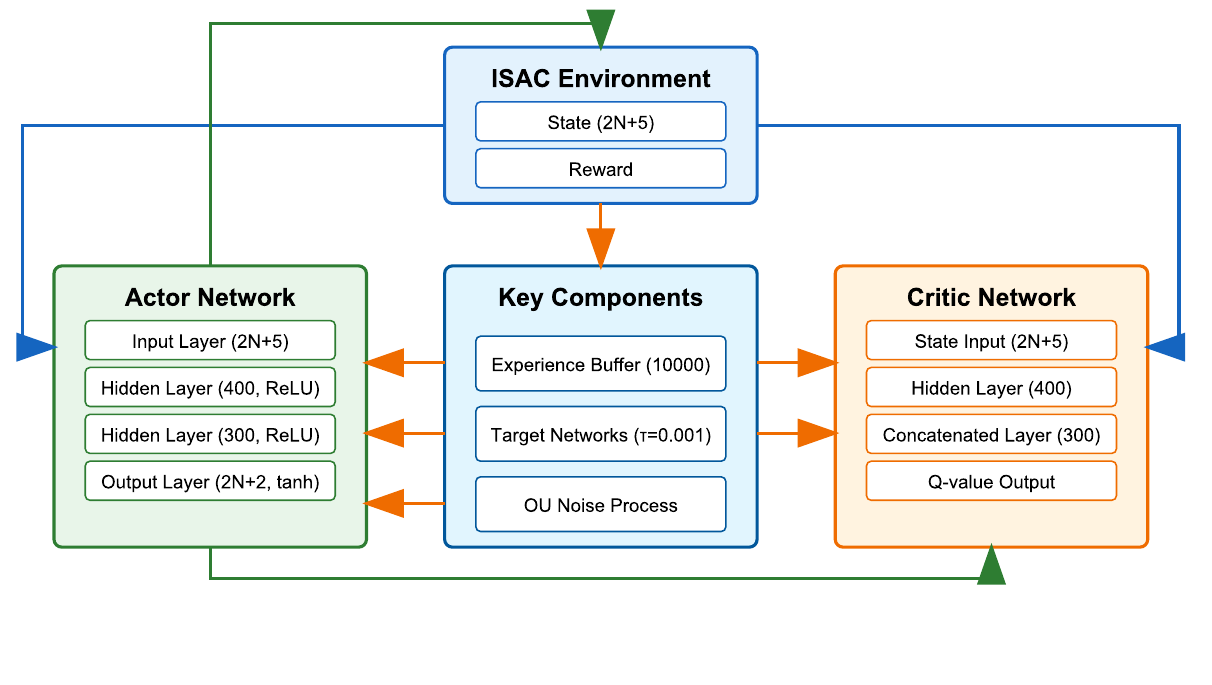} \vspace{-4mm}
		\caption {The framework of DDPG.}
		\label{ddpg} \vspace{-3mm}
	\end{figure}

	\subsubsection{DDPG Framework}
	As illustrated in Fig. \ref{ddpg}, the design and training strategy of the DDPG framework are detailed in the following paragraphs.
	
	\paragraph{Actor Network}
	The Actor network $\mu(s|\theta^\mu)$ is constructed with three fully connected layers. The input layer receives the state vector of dimension $2(N+1)+3$, which contains both antenna positions and beamforming features. The hidden layers consist of 400 and 300 neurons respectively with ReLU activation functions to extract hierarchical features. The output layer is activated by tanh function and scaled by the action bound $A/4$. 
	
	\paragraph{Critic Network}
	The Critic network $Q(s,a|\theta^Q)$ is designed to estimate the action-value function and guide the Actor's policy update. It processes the state through a fully connected layer of 400 neurons, after which the action is concatenated with the processed state features. This concatenated representation then goes through another fully connected layer of 300 neurons with ReLU activation. Finally, a single output neuron predicts the Q-value. 
	
	\paragraph{Training Strategy}
	In RL with continuous actions, effective exploration is crucial yet challenging. We employ an Ornstein-Uhlenbeck (OU) process rather than simple Gaussian noise to add temporally correlated exploration noise to the actor's actions. This enables smooth and consistent FAs' movements \footnote{The OU process generates mean-reverting noise patterns suitable for physical control tasks.}.  The OU  process is characterized by 
	\begin{equation}\label{eq15}
		\mathcal{Z}_{t+1} = \mathcal{Z}_t + \xi(0-\mathcal{Z}_t) + \varsigma\mathcal{N}(0,1),
	\end{equation}
	where $\xi$ controls mean reversion and $\varsigma $ determines noise magnitude. Furthermore, We maintain target networks $\mu'$ and $Q'$ with parameters $\theta^{\mu'}$ and $\theta^{Q'}$, updated via the following equation: \vspace{-2mm}
	\begin{equation}\label{eq16}
		\theta' \leftarrow \tau\theta + (1-\tau)\theta',
	\end{equation} 
	where $\tau = 0.001$ ensures stable learning. Besides, a buffer of size 10000 stores transitions $(s_t,a_t,r_t,s_{t+1})$, from which mini-batches of size 64 are sampled for network updates.

	\begin{figure}[t]
		\hspace{-3mm}\centering
		\includegraphics[width=2.8in]{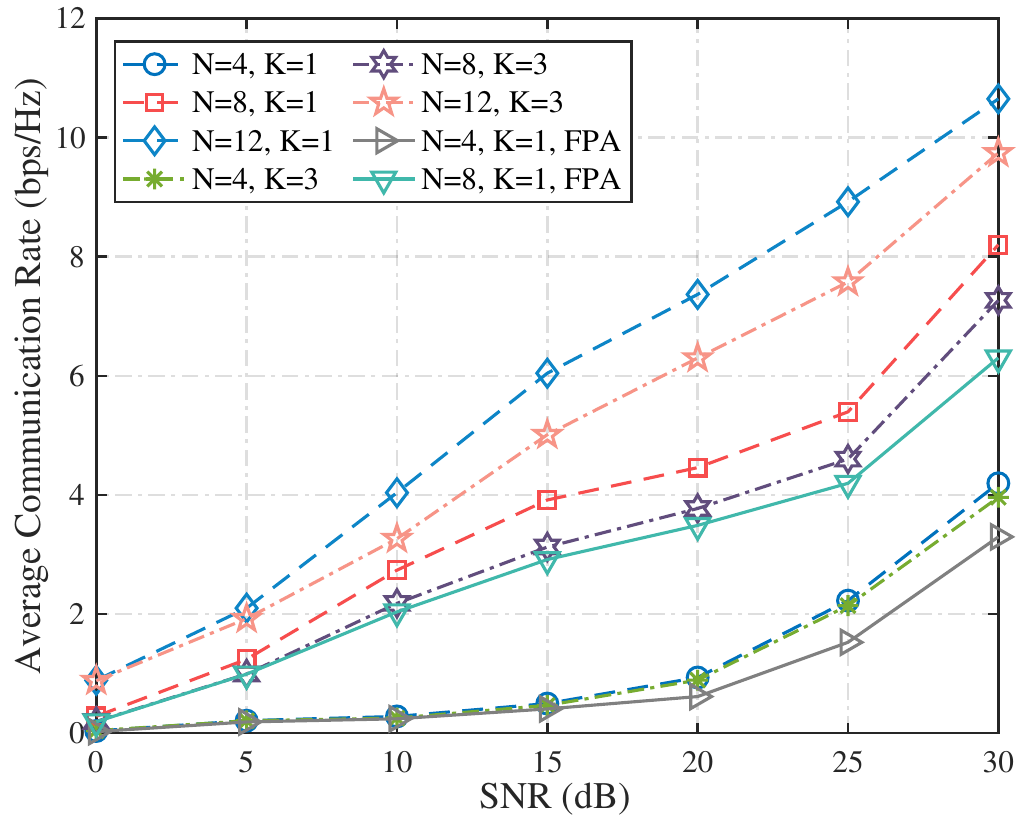}
		\caption {Average Communication rate.}
		\label{ave}
	\end{figure}  
	
			\begin{figure}[t]
		\centering
		\includegraphics[width=3.3in]{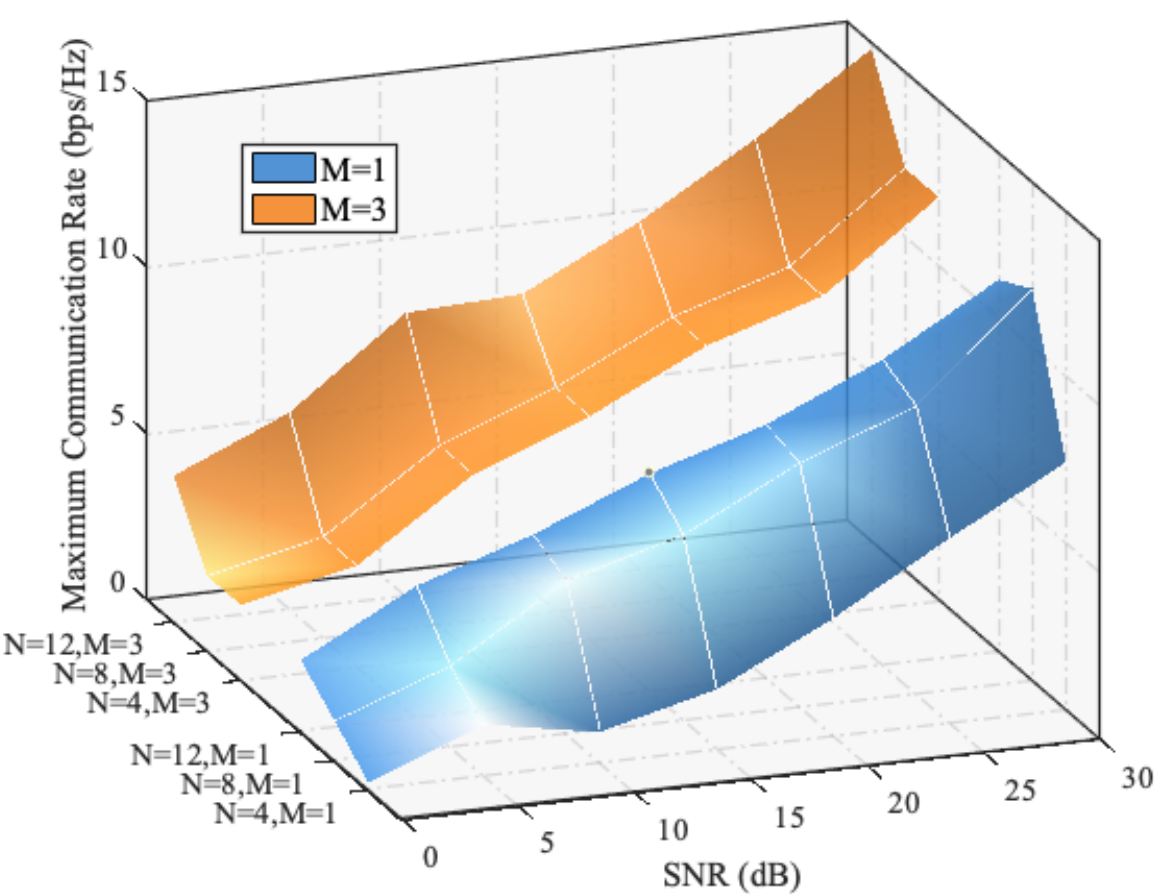}
		\caption {Maximum Communication rate.}
		\label{max}
	\end{figure}

	\section{Numerical Results} \vspace{-1mm}
	In our simulation experiments, we consider that the elevation and azimuth angles $\theta^{(t)}_{k}, \psi^{(t)}_{k}, \theta^{(r)}_i, \psi^{(r)}_i$ are all independent and identically distributed variables randomly distributed in $[0,\pi]$. The minimum distance constraint between the fluid antennas is set to $D=\lambda/2$ and restricted to movement within the range of $A \times A$. The path response matrix is assumed to be diagonal with
	$\mathbf{\Sigma}[1,1]\sim \mathcal{CN}({0}, {\tau/(\tau+1)})$ and $\mathbf{\Sigma}[d,d]\sim \mathcal{CN}({0}, {1/(\tau+1)(D-1)})$ for $d=2,3,\dots,D$, where $\tau=1$ represents the ratio of the average power of the line-of-sight (LoS) path to the average power of the non-line-of-sight (NLoS) path. We assume that the number of transmit and receive paths as $D=I=3$.

Fig.~\ref{ave} illustrates the average communication rate of the proposed FAS-ISAC system under varying numbers of targets \( K \). As \( K \) increases, the ISAC constraints become more stringent, leading to a reduction in communication rates. Nevertheless, the FAS consistently outperforms the fixed-position antenna (FPA) baseline. Even with multiple targets (\( K > 1 \)), the communication rate achieved by the FAS remains higher than that of the FPA with only one target. \textit{This result highlights the superior performance of the FAS, effectively balancing communication and sensing tasks despite the increased sensing demand.}

Fig.~\ref{max} focuses on the maximum achievable communication rate, demonstrating the impact of increasing the number of UT \( M \). With the addition of more UT, the system achieves a notable improvement in maximum rates. For example, at 30 dB SNR with \( N = 12 \) antennas, the \( M=3 \) configuration achieves a maximum communication rate of 14.84 bps/Hz, compared to 11.64 bps/Hz for the \( M=1\) configuration, an improvement of approximately $27.6\%$. This highlights the potential of multi-user scenarios to unlock higher system performance. Furthermore, the proposed DRL framework seamlessly adapts to multi-user setups, intelligently positioning antennas to optimize both communication and sensing objectives. \textit{These results validate the scalability and robustness of the DRL framework, demonstrating its effectiveness in handling complex, multi-user environments while enabling real-time optimization.}

	\section{Conclusion} 
	This letter proposed a  BCD-DRL based approach for intelligent antenna positioning in FAS-aided ISAC systems, which is well suited to handle multiple targets. By leveraging  DDPG algorithm, the framework effectively addressed the joint optimization of beamforming and antenna positions, balancing sensing and communication performance. Simulation results demonstrated the scalability and efficiency of FAS and the proposed approach, providing valuable insights for practical deployment of FAS-ISAC systems.

\end{document}